\documentclass[traditabstract]{aa}

\def\apjl{ApJ}%
\def\apjs{ApJS}%
\def\ao{Appl.~Opt.}%
\def\aap{A\&A}%
\usepackage{graphicx,amssymb,natbib}

\title{Weak lensing observations of potentially X-ray underluminous
  galaxy clusters} 
\author{J.\,P. Dietrich\inst{1}
\and
A. Biviano\inst{2}
\and
P. Popesso\inst{3}
\and 
Y.-Y. Zhang\inst{4}
\and
M. Lombardi\inst{1}
\and
H. B\"ohringer\inst{3}}
\institute{ESO, Karl-Schwarzschild-Str. 2, 85748 Garching
  b. M\"unchen, Germany\\
\email{jdietric@eso.org}
\and
INAF-Osservatorio Astronomico di Trieste, via G. B. Tiepolo 11,
I-34143 Trieste, Italy 
\and
Max-Planck Institut f\"ur extraterrestrische Physik,
Gie{\ss}enbachstr., 85748 Garching b. M\"unchen, Germany
\and
Argelander-Institut f\"ur Astronomie, Auf dem H\"ugel 71, 53121 Bonn,
Germany}

\begin{document}

\date{Received $t_0$ / Accepted $t_1 > t_0$}

\abstract{Optically selected clusters of galaxies display a relation
  between their optical mass estimates and their X-ray luminosities
  $L_\mathrm{X}$ that has a large scatter. A substantial fraction of
  optically selected clusters have $L_\mathrm{X}$ estimates or upper
  limits significantly below the values expected from the
  $L_\mathrm{X}$-mass relation established for X-ray selected
  clusters, i.e., these clusters are X-ray underluminous for their
  mass.
  We attempt to confirm or falsify the X-ray underluminous nature of
  two clusters, Abell~315 and Abell~1456, by using weak gravitational
  lensing as a third and independent measure of the clusters' masses.
  We obtained optical wide-field imaging data and selected background
  galaxies using their colors and measured the shear exerted by the
  tidal field of the foreground galaxy clusters. We then fitted
  parametrized models to our shear catalogs.
  After accounting for projections of large-scale structure and halo
  triaxiality, we find that A~315 is significantly X-ray underluminous
  for its mass, while no significant lensing signal was detected for
  A~1456. We re-evaluate earlier kinematic and X-ray analyses of these
  two clusters and discuss the nature of the X-ray underluminous
  cluster A~315 and why A~1456 was probably erroneously identified as
  being X-ray underluminous.

}

\maketitle

\section{Introduction}
\label{sec:introduction}
Clusters of galaxies can be detected by various methods: optical
\citep[e.g.,][]{2006astro.ph..1195G}, X-ray emission
\citep[e.g.,][]{2001A&A...369..826B}, weak gravitational lensing
\citep[e.g.,][]{1996MNRAS.283..837S,2007A&A...470..821D}, and the
Sunyaev-Zeldovich effect \citep[e.g.,][]{2008arXiv0810.1578S}.
Differences in the selection method could potentially lead to biases
when determining the cluster mass function. Optical selection is
generally more affected by projection effects than X-ray selection,
although projection effects can be minimized by selecting cluster
galaxies on the basis of their colors
\citep[e.g.,][]{2002AJ....123...20K}. On the other hand, X-ray
selection requires that the intra-cluster gas has been heated to a
detectable level, and theoretical predictions show that there is a
non-negligible fraction of unvirialized, relatively massive clusters
with no X-ray emission \citep{2002MNRAS.337.1269W}. While such systems
are mostly expected at high redshifts, they may be present at all
epochs.

Most cluster mass functions have been obtained from X-ray selected
samples \citep[e.g.,][]{2002ApJ...567..716R}. X-ray selection is
generally considered to be well understood, almost pure and complete
in mass. However, several investigations have questioned the
completeness of X-ray selected cluster samples. These investigations
have shown that there is a population of \emph{optically selected}
clusters that deviate from the relation between the X-ray luminosity,
$L_\mathrm{X}$, and virial mass (or an optical mass proxy such as
richness and velocity dispersion) established for X-ray selected
cluster samples \defcitealias{2007A&A...461..397P}{P07}\citep[P07
hereinafter]{1997MNRAS.291..353B,2001ApJ...558..590M,2002ApJ...569..689D,2004MNRAS.348..551G,2004ApJ...601L...9L,2006ApJ...645..955B,2007ApJ...660L..27F,2007A&A...461..397P}.
These clusters are underluminous in X-ray for their masses. P07, in
particular, identified X-ray underluminous clusters in the RASS-SDSS
survey among Abell clusters, which were therefore named ``Abell X-ray
Underluminous'' (AXU). The virial masses of these AXU clusters were
determined from the redshifts of their member galaxies, using data
from the Sloan Digital Sky Survey. X-ray luminosities or upper limits
to X-ray luminosities were derived from the ROSAT All-Sky Survey.

What is the nature of these AXU clusters?  By analyzing the velocity
distribution of their members, \citetalias{2007A&A...461..397P}
suggested they are systems in the process of formation, . However, it
is possible that at least part of the scatter in the
$L_\mathrm{X}$-mass relation is not intrinsic, but originates in
erroneous estimates of either the cluster $L_\mathrm{X}$ or its mass.
To investigate this point, we obtained optical wide-field observations
of two AXU clusters with the aim of measuring these clusters' masses
by means of weak lensing, and deriving a third, independent mass
estimate. We describe these data and their reduction in
Sect.~\ref{sec:data}. Section~\ref{sec:weak-lens-mass} contains the
weak lensing analysis, which we compare with revised kinematic and
X-ray estimates in Sects.~\ref{sec:kinematic-mass} and~\ref{sec:xray}.
We revisit the optical cluster luminosity and richness in
Sect.~\ref{sec:optical} and present our conclusions in
Sect.~\ref{sec:discussion}.

Throughout this paper, we use a standard $\Lambda$CDM cosmology with
$\Omega_\mathrm{m} = 0.3$, $\Omega_\Lambda = 0.7$, and $H_0 =
70\,h$\,km\,s$^{-1}$\,Mpc$^{-1}$. Confidence intervals correspond to
the 68\% confidence level.

\section{Data}
\label{sec:data}
\object{Abell~315}, a galaxy cluster at $z=0.174$
\citepalias{2007A&A...461..397P}, was observed with the Wide-Field
Imager (WFI) at the ESO/MPG-2.2\,m telescope in B-, V-, and R-band.
WFI is a focal-reducer type camera with a $4 \times 2$ mosaic of
$2\mathrm{k} \times 4\mathrm{k}$ CCDs with a filling factor of
$95.9\%$ \citep{1999Msngr..95...15B}. Its field-of-view (FOV) is
$34\arcmin \times 33\arcmin$, resulting in a pixel scale of
$0\farcs238$. At the cluster redshift, $1\arcmin$ corresponds to a
physical scale of $177\,h^{-1}$\,Mpc. The observations were carried
out in service mode during the nights from Nov. 5 to 12, 2007 in dark and
clear sky conditions. The total exposure times for the three bands were
2880\,s, 5890\,s, and 5500\,s, respectively.

\object{Abell~1456}, which is at a redshift of $z=0.135$
\citepalias{2007A&A...461..397P}, was also observed with WFI in the
same passbands. For A~1456, $1\arcmin$ corresponds to
$144\,h^{-1}$\,Mpc. These observations were performed during the
nights of May 9 and 10, 2008 in dark and clear conditions. The total
exposure times for the A~1456 field were 1500\,s, 5750\,s, and 5500\,s
for the B-, V-, and R-band, respectively.

All data were processed with the GaBoDS/THELI pipeline
\citep{2005AN....326..432E}. Because no photometric standard stars
were observed in the nights that our data were taken, the
transformation equations of \citet{2004AN....325..299K} were used to
calibrate the V- and R-band data using SDSS magnitudes of objects in
the same fields. The zero points of the B-band data were fixed by
matching the expected stellar colors of the
\citet{1998PASP..110..863P} stellar library in a color-color diagram
to the observed colors of stars in the fields. The colors of galaxies
were corrected for Galactic extinction using the
\citet{1998ApJ...500..525S} extinction maps. The effective seeing of
the coadded images R-band images is $0\farcs74$ (A~315) and
$1\farcs0$ (A~1456).

The observations were divided into observing blocks (OBs) of 5
dithered exposures. Additional offsets between OBs ensured that the
sky coverage of the observations was more homogeneous. Every region in
the final V- and R-band images was covered by at least six exposures.
In the shorter B-band observations, which used only one OB, every
location was covered by at least three exposures. The WFI
point-spread-function (PSF) is smooth across chip gaps and has only
slow spatial variations, so that the coaddition of dithered exposures
does not lead to discontinues in the PSF within the final image.

\begin{figure*}
\includegraphics[width=\textwidth]{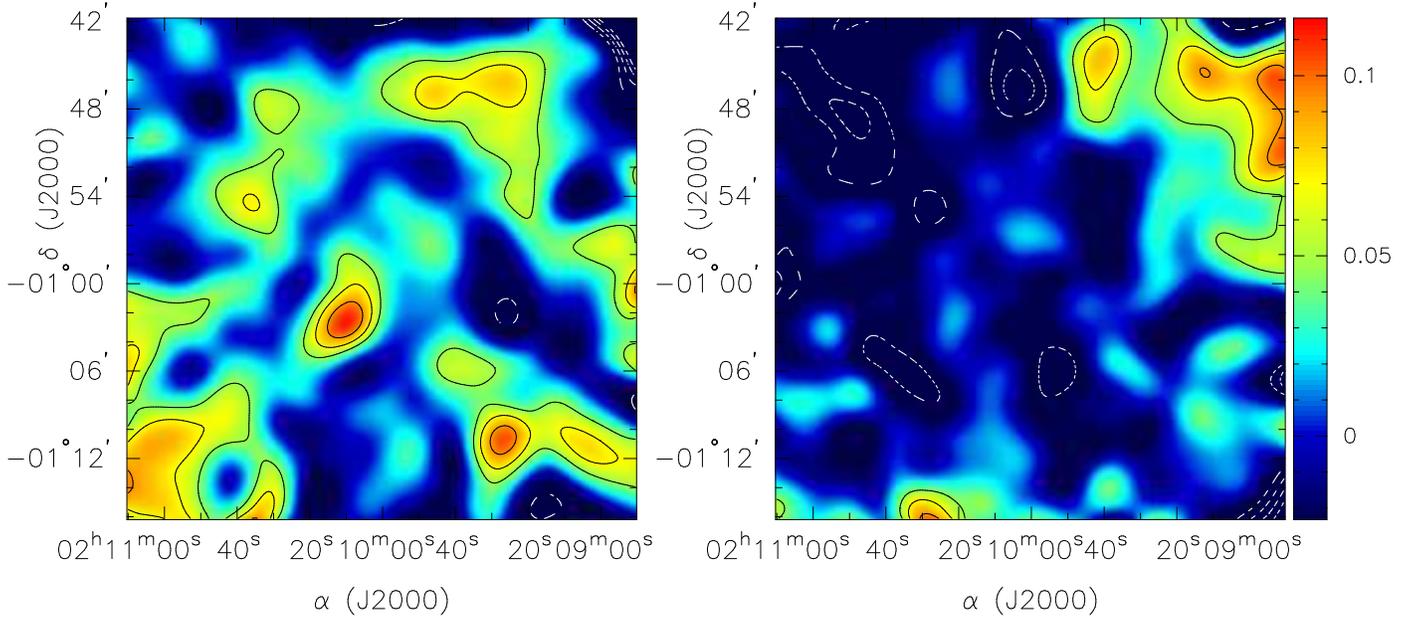}
\caption{\emph{Left}: Mass reconstruction of the A~315 field. The
  surface mass density contours start at at 0.0464 (corresponding to
  $1.6 \times 10^{14}\,h\,M_\odot$\,Mpc$^{-2}$ or $2\sigma$ above the
  mean surface mass density at the edge of the field),
  increasing in steps of $1\sigma$. Dashed contours are at the
  same negative levels. \emph{Right}: B-mode reconstruction of the
  A~315 field. Contours are at the same levels as in the left panel.}
  \label{fig:a315_rec}
\end{figure*}

\begin{figure*}
\includegraphics[width=\textwidth]{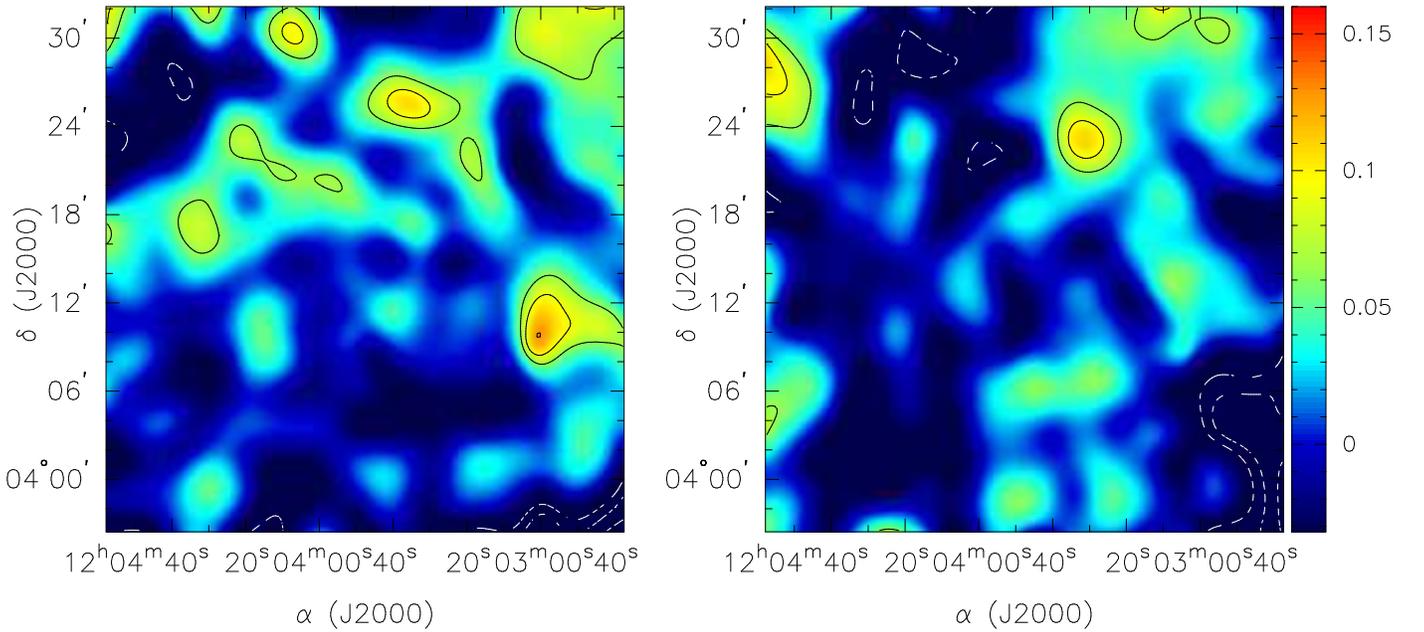}
\caption{Same as Fig.~\ref{fig:a315_rec} for the A~1456 field. The
  $1\sigma$ level is at 0.032 (corresponding to $1.25 \times
  10^{14}\,h\,M_\odot$\,Mpc$^{-2}$). The color bar on the right runs
  from $-1\sigma$ to $5\sigma$ in both figures.}
  \label{fig:a1456_rec}
\end{figure*}

\subsection{Lensing catalogs}
\label{sec:shape-measurement}

The Bayesian shape fitting method \emph{lens}fit\footnote{If the
  reader wishes to use the \emph{lens}fit code, its authors can be
  contacted at \texttt{http://www.physics.ox.ac.uk/lensfit/}}
\citep{2007MNRAS.382..315M,2008arXiv0802.1528K} was used to estimate
the shear signal imprinted on the shapes of background galaxies. Stars
were pre-selected in a magnitude-flux-radius diagram with the
additional requirement that \texttt{SExtractor}'s CLASS\_STAR $>0.95$.
These stars were modeled with the elliptical Gauss-Laguerre method
\citep{2002AJ....123..583B,2007AJ....133.1763N} and a two dimensional
4th order polynomial was fitted to the Gauss-Laguerre coefficients to
obtain a PSF model over the entire WFI field-of-view. \emph{Lens}fit
-- in the implementation used here -- subdivides the image area into
regular grid cells, inside which the PSF is assumed to be constant.
Here the side length of the cells was chosen to be 100 pixels or
$0\farcm4$. For each cell, an image of the PSF was reconstructed from
the shapelet coefficients at the cell's center. These PSF images were
then used as a PSF description for \emph{lens}fit.

The Bayesian prior function proposed by \citet{2008arXiv0802.1528K},
\begin{eqnarray}
  \label{eq:1}
  \mathcal{P}(e_1, e_2) = A \cos\left(\frac{|e|\pi}{2}\right)
  \exp\left[ -\left(\frac{2|e|}{B (1+|e|)^D)}\right)^C\right]\;,
\end{eqnarray}
where $B$, $C$, and $D$ are free parameters and $|e| =
\sqrt{e_1^2+e_2^2}$, was used in computing the posterior ellipticity
probability distribution. Because of the many degeneracies of this
function, a brute-force minimization was the only viable option in
determining its coefficients. The coefficients found here are very
similar to those obtained by \citet{2008arXiv0802.1528K} for the STEP1
simulations \citep[][see Table~\ref{tab:prior} for a
comparison]{2006MNRAS.368.1323H}.

\begin{table}
  \caption{Coefficients of the lensfit prior function.}
  \centering
  \begin{tabular}{lrrr}
    \hline\hline
    Coeff. & STEP1 & A~315 & A~1456  \\\hline
    B & $0.029$ & $0.042 $ & $0.068$ \\
    C & $0.45$  & $0.44$   & $0.44$  \\
    D & $42.7$  & $32.6$   & $30.5$  \\\hline
  \end{tabular}
  \label{tab:prior}
\end{table}

Shear estimates for individual galaxies were obtained using Eq.~(20)
of \citet{2008arXiv0802.1528K}. When measuring the shear we must -- as
in Eq.~(\ref{eq:1}) -- assume a prior that contains zero shear because
we cannot predict the variation in shear across the FOV. Since after
lensing the ellipticity distribution of galaxies is not centered on
zero anymore, we need to apply an additional weighting factor to
correct for the assumption of zero shear. The shear sensitivity
$\partial e_i/\partial g_i$ describes how the measured ellipticity $e$
of a galaxy depends on the reduced shear $g$. Very low signal-to-noise
ratio (SNR) objects have sensitivity values close to zero. Together
with high ellipticity measurements, this can lead to unphysically high
shear values. Therefore, shear estimates with shear sensitivity of
$\partial e_i/\partial g_i < 0.3$ or ellipticity $|e| > 1.1$ were
rejected. We note that both types of rejection to low SNR objects
affected fewer than 50 galaxies, and hence did not bias our lensing
estimates.

Because this is the first application of \emph{lens}fit to real data,
we compared the shear estimates obtained in this way with those
computed using the \citet{2001A&A...366..717E} implementation of the
\citet[][KSB]{1995ApJ...449..460K} shape estimation algorithm. The
STEP1 bias parameters $q$ (non-linear response), $m$ (calibration
bias), and $c$ (constant offset) were computed where the
\emph{lens}fit shear estimates were set to $\gamma_i^\mathrm{true}$.
We found that $q$ and $c$ are consistent with zero and that
$\overline{m} = 0.152\pm0.029$. This is consistent with the bias
parameter values of the MH implementation of KSB in
\citet{2006MNRAS.368.1323H}, which is, save a few details, identical
to the KSB implementation used here. This result suggests that, apart
from errors in the PSF model common to both methods, the
\emph{lens}fit shear estimates are very close to the true shear
values.

Background galaxies were selected based on their magnitude and colors.
Galaxies with $R>23$\,mag were considered to be background galaxies
and were included in the catalog irrespective of their colors, while
brighter galaxies were rejected if their colors matched the colors of
cluster galaxies, $0.53 < (B-V) < 1.38$ and $-0.25 < 1.7 \times (V-R)
- (B-V) < 0.3$. Finally, only galaxies manually preselected in a
magnitude--flux-radius diagram were kept if their flux radius was
$<1\farcs3$. The effectiveness of this color selection was tested by
plotting the number density of galaxies in radial bins from the
cluster centers. A small excess of galaxies was found only for A~315
within a clustercentric distance of $1\farcm2$; beyond this radius,
the number density remained constant, which is indicative of no
contamination of the background catalogs by cluster galaxies.

After these selections, $9505$ galaxies with shear estimates remained
in the lensing catalog of A~315, and $8030$ galaxies were left in the
A~1456 catalog, corresponding to respective number densities of
$8.6$\,arcmin$^{-2}$ and $7.5$\,arcmin$^{-2}$, where areas masked due
to the occurrence of bright stars, reflection rings, and diffraction
spikes were excluded.

\section{Weak lensing mass estimates}
\label{sec:weak-lens-mass}
Based on the lensing catalogs described in the previous section,
surface mass density maps were computed using the finite field
inversion method of \citet{2001A&A...374..740S} with a smoothing
scale of $2\arcmin$. These maps are presented in the left panels of
Figs.~\ref{fig:a315_rec} and~\ref{fig:a1456_rec}. The right panels of
these figures are B-mode maps obtained from shear catalogs with
galaxies rotated by $45\degr$ to cancel any true shear signal,
and is a test of systematic residuals in the PSF correction. The
flatness of the B-mode maps suggests that the PSF models used in
creating the lensing catalogs are sufficiently accurate for obtaining
cluster mass estimates. The noise levels of the reconstructions were
estimated from the variance of 100 maps with randomized galaxy
orientations.

At a confidence level just under $5\sigma$, A~315 is clearly detected
a little SE of the image center. No lensing signal is visible at the
position of A~1456 at the center of the image. The strong peak at the
right of the image is probably caused by the reflection ring of a
bright star on which it is centered. However, this peak is also only
$45\arcsec$ away from the cluster candidate NSC\,J120257+040951
\citep{2003AJ....125.2064G}.

The weak lensing peak of A~315 is centered on a galaxy concentration
whose brightest galaxy is $2\farcm2$ from the cluster position of
\citet{1989ApJS...70....1A} and $4\farcm3$ from the cluster center
adopted in the X-ray and kinematic analysis of
\citetalias{2007A&A...461..397P}. Both the position of
\citetalias{2007A&A...461..397P} and the brightest cluster galaxy
(BCG) identified here (02:10:06.46, $-$01:01:56.46), are compatible
with the positional uncertainties of \citet{1989ApJS...70....1A} but
are mutually incompatible. Because the
\citetalias{2007A&A...461..397P} position of A~315 was based on a
very weak X-ray signal, the BCG position was adopted as the cluster
center for the analysis presented here, a choice that we further justify
in Sect.~\ref{sec:kinematic-mass}, where the kinematic analysis of
both clusters is re-examined. For A~1456 the weak lensing analysis was
done taking the BCG identified in Sect.~\ref{sec:kinematic-mass}
(12:03:48.71, +04:20:43.46), located $5\farcm75$ from the X-ray center
of \citetalias{2007A&A...461..397P}, to be the center.

Converting the dimensionless lensing quantities into physical mass
densities requires knowledge of the source redshift distribution or at
least of the average source redshift. To determine the latter
quantity, galaxies were randomly drawn from the photometric redshift
catalog of \citet{2006A&A...457..841I} to match the magnitude
distribution of our lensing catalogs, using the color transformations
of \citet{2007AJ....133..734B}. The average redshifts of the shear
catalogs are $\overline{z}_\mathrm{A 315} = 1.07$ and
$\overline{z}_\mathrm{A 1456} = 1.03$. While it is formally possible
to estimate errors in these mean redshifts by bootstrap resampling
from the CFHTLS catalogs, the true redshift error will be dominated by
cosmic variance, which we do not explicitly calculate here. We note
that for the two low-redshift clusters under investigation, a
redshift error as high as $0.2$ would result in a mass error of less
than $5\%$ and even in this extreme case would be an insignificant
contribution to the total error budget of the lensing mass estimates.

To obtain mass estimates, parametrized models -- singular isothermal
spheres (SIS) and NFW profiles -- were fitted using the maximum
likelihood method of \citet{2000A&A...353...41S}. The best-fit SIS
model for A~315 has a velocity dispersion of $\sigma_\mathrm{SIS} =
747^{+85}_{-82}$\,km\,s$^{-1}$. The best-fit NFW model for A~315 has
$M_{200} = 2.96^{+1.15}_{-0.75} \times 10^{14}\,h^{-1}\,M_\odot$. The
minimization was carried out with a Downhill Simplex method
\citep[e.g.,][]{1992nrca.book.....P} in which at every vertex the
concentration parameter was fixed to the prescription of
\citet{2004A&A...416..853D}. Attempts to fit parametrized models to
A~1456 returned only marginally significant results. The best-fit SIS
has $\sigma_\mathrm{SIS} = 418^{+150}_{-257}$\,km\,s$^{-1}$, and the
NFW model has $M_{200} = 6.3^{+7.0}_{-5.1} \times
10^{13}\,h^{-1}\,M_\odot$. All model fits were centered on the BCG.

\citet{2001A&A...370..743H,2003MNRAS.339.1155H} and
\citet{2004PhRvD..70b3008D} studied the influence of uncorrelated
large-scale structure (LSS) on cluster mass estimates.
\citet{2004PhRvD..70b3008D} found that LSS projections can increase
the error in $M_{200}$ by as much as $75\%$ when the concentration
parameter $c$ and $M_{200}$ are fitted simultaneously. Much of this
additional error comes from the stretching of the error ellipsis along
the $c$-axis, and it was estimated from Fig.~3 of
\citet{2004PhRvD..70b3008D} that in the case of fixed $c$, as used in
this work, the additional error is at most $50\%$.

\citet{2007MNRAS.380..149C} investigated the influence of halo
triaxiality on weak-lensing mass estimates and found that very oblate
or prolate halos with axis ratio 1:3 decrease or increase weak-lensing
mass estimates by as much as $40\%$, depending on their orientation
with respect to the line of sight. These are, however, extreme cases,
which occur in less than $1\%$ of all halos
\citep{2005ApJ...629..781K} and a realistic estimation of the
contribution to the total error budget must take the halo-shape
distribution into account. \citet{2006ApJ...646..815S}, for example,
found that prolate halos, i.e., those halos that according to
\citet{2007MNRAS.380..149C} lead to higher additional errors in weak
lensing measurements, are much more common than oblate halos.
\citet{2005ApJ...629..781K} presented a fitting function for the
minor-axis as a function of mass and redshift. For a halo at the
redshift and mass of A~315 the expected $c/a = 0.63$. From Figs.~4
and~5 of \citet{2007MNRAS.380..149C}, we found that in the extreme
cases of the major (minor) axis being aligned with the line of sight,
the weak-lensing mass for this axis ratio will be overestimated
(underestimated) by $16\%$ ($10\%$). Because the results of
\citet{2007MNRAS.380..149C} cannot be convolved with the full
halo-shape distribution, we adopt this combination of mean axis ratio
and extreme alignment as additional error from halo triaxiality.

Adding the systematic errors of LSS projections and halo triaxiality
leads to total errors that are significantly higher than those
obtained from the model fits alone. For A~315 $M_{200} =
(2.96^{+1.15\, +0.69}_{-0.75\, -0.50}) \times
10^{14}\,h^{-1}\,M_\odot$ was found, where the first error is random
and the second error is sytematic. The marginal detection of A~1456
turns into an upper limit to the mass if projection effects of LSS
are taken into account, $M_{200} = (6.3^{+7.0\, +4.2}_{-5.1\, -3.4})
\times 10^{13}\,h^{-1}\,M_\odot$.

\section{Kinematical mass estimates}
\label{sec:kinematic-mass}

\begin{figure}
  \resizebox{\hsize}{!}{\includegraphics[clip]{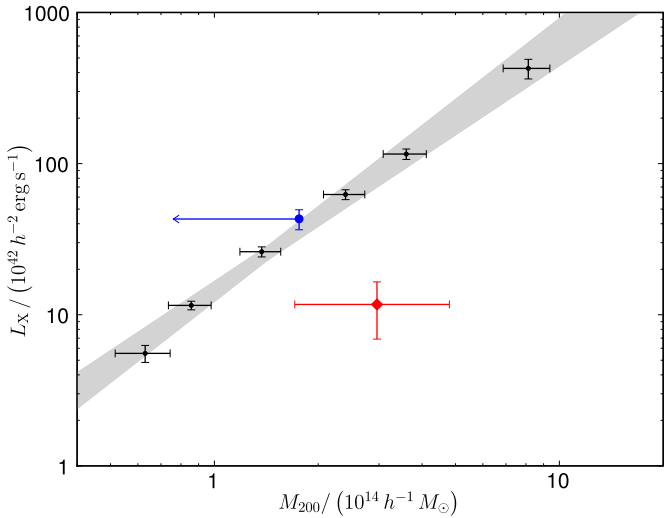}}
  \caption{The clusters A~315 (red diamond) and A~1456 (blue circle)
    in comparison with the $L_\mathrm{X}$-$M_{200}$ relation of
    \citet[black dots and shaded area]{2008MNRAS.387L..28R}. The error
    bars for the mass of A~315 are the sum of random and systematic
    errors. For A1456, we use the $1\sigma$ random plus systematic
    upper limit to the mass estimate, since the cluster is undetected
    in the weak lensing analysis.}
  \label{fig:m-lx-relation}
\end{figure}

\begin{figure}
  \resizebox{\hsize}{!}{\includegraphics{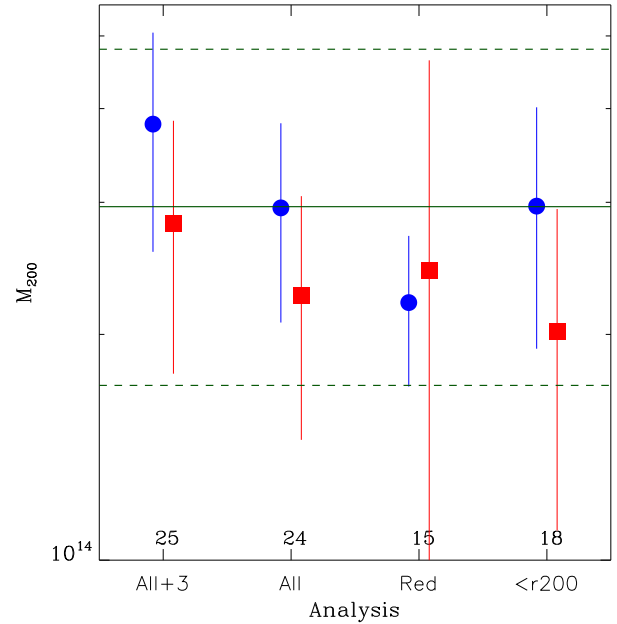}}
  \caption{Comparison of kinematic and weak-lensing mass estimates of
    A~315. The filled (blue) circles are virial mass estimates; the
    filled (red) squares are mass estimates obtained from the velocity
    dispersion. Shown are the estimates for the samples of (1) all
    galaxies plus the 3 outliers (see text for details); (2) all
    galaxies; (3) only red-sequence galaxies; and (4) galaxies within
    $r_{200}$. The numbers on the $x$-axis indicate the sample sizes.
    The solid (green) line shows the weak-lensing mass estimate with
    error bars (dashed lines) for comparison.}
  \label{fig:a315mass}
\end{figure}

The lensing analysis in the previous section established that the
position of A~315 is differs significantly from the one used in the
kinematical analysis of \citetalias{2007A&A...461..397P}. Because the
choice of centroid position affects the selection of cluster members,
the kinematic analysis of SDSS DR5 data \citep{2007ApJS..172..634A}
was repeated using the BCG as the cluster center. This analysis
follows the steps of \citet{2006A&A...456...23B}, as briefly
summarized here:

\begin{enumerate}
\item a peak in the redshift distribution is selected;
\item interlopers are rejected by analyzing the projected phase-space
  diagram, i.e., velocities versus clustercentric distances;
\item the virial mass and velocity dispersion are computed within a
  given aperture;
\item the computed mass and velocity dispersion are translated into
  the mass within $r_{200}$.
\end{enumerate}
The mass estimates are corrected for the surface-pressure term
\citep{1986AJ.....92.1248T}. In the following, we describe the
influence of these steps on the kinematical mass estimates of A~315.

Step 1 requires the choice of an aperture centered on the adopted
center position to select all galaxies within this aperture. Weighted
gap \citep{1990AJ....100...32B} and the density gap
\citep{1998A&A...331..493A} estimators with a gap value of $4$ were
chosen to select the peak position. We note that our mass estimates
are insensitive to the choice of the gapper method and of the gap
value. However, the mass value depends on the initial aperture for the
search of the peak. Adopting a radius of $2\,h^{-1}$\,Mpc instead of
$1\,h^{-1}$\,Mpc causes the mass estimate of A~315 to increase by
$31\%$. This is because of the inclusion of 3 galaxies in the data-set
that are excluded when an aperture of $1\,h^{-1}$\,Mpc is used. With
the exception of these 3 galaxies, the identification of interlopers
(step 2) is quite robust, with little room for a different selection.

The mass estimate in step 3 may be computed within an aperture that 
differs from the one used in step 1 by being larger. Here
one Abell radius, $2.15\,h^{-1}$\,Mpc, was adopted. The mass within
this radius is used to obtain a first estimate of $r_{200}$, and then
to interpolate from the Abell radius to $r_{200}$ using an NFW profile
and obtain $M_{200}$ (step 4). Alternatively, instead of using the
Abell radius, the mass within the aperture corresponding to $r_{200}$
can be computed. Since the $r_{200}$ estimate will change as a result
of the aperture selection, this is done iteratively. The difference
between the mass estimates of both methods is negligible.

Rather than using all cluster galaxies, one can use only those on the
red sequence. In this case, we found that the mass estimate decreases
by $25\%$, if the mass estimated is obtained within an aperture of
$2.15\,h^{-1}$\,Mpc. 

Two mass estimates are reported at the end of each analysis. One comes
from the virial analysis and the other uses the velocity dispersion
as a proxy. It is known from numerical simulations
\citep{2006A&A...456...23B} that the former is generally an
overestimate and the latter an underestimate, so that the two values
should bracket the real one.

Mass errors were computed with the Jackknife technique. They range
from $23\%$ to $35\%$ for the virial mass, but are much higher for the
mass derived from the velocity dispersion (from $36\%$ to $90\%$). The
various mass estimates of A~315 are displayed in
Fig.~\ref{fig:a315mass}. All mass estimates reported here are lower
than the value of $(6.61 \pm 2.71) \times 10^{14}\,h^{-1}\,M_\odot$
reported by \citetalias{2007A&A...461..397P}. The difference is due to
the different central position adopted here, the BCG rather than the
X-ray center. This centroid choice also decreases the systematic error
estimates because the main peak seems to be more reliably centered.

We consider the range in the estimates obtained by changing the
parameters of the analysis to be systematic error, and the average
jackknife errors of these estimates as random error. The cluster mass
is now estimated to be $(2.7^{+1.1}_{-0.7}\pm 1.0) \times
10^{14}\,h^{-1}\,M_\odot$, where the first error is the random
error and the second one is systematic.

According to a \citet{1988AJ.....95..985D} test, there is no evidence
of subclustering, but this test is not very reliable for samples of
redshift for fewer than 50 galaxies; there is also no evidence of a
gradient in the $R$--$v$ plane.

\begin{figure}
  \resizebox{\hsize}{!}{\includegraphics{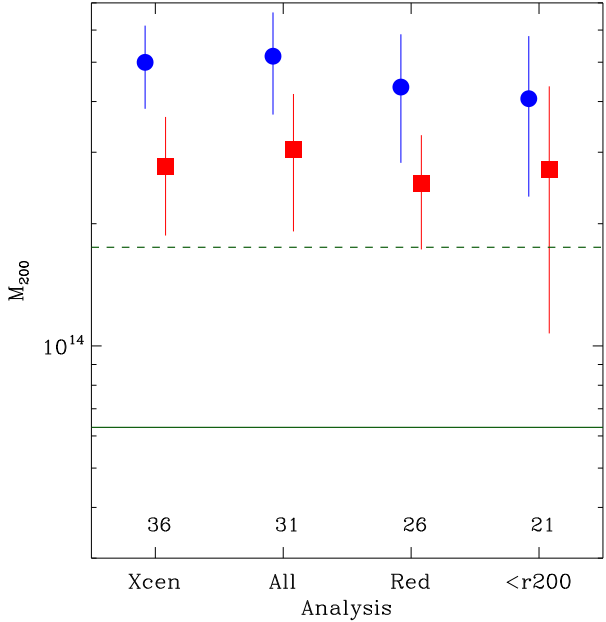}}
  \caption{Same as Fig.~\ref{fig:a315mass} but for A~1456. The
    leftmost point is for the analysis performed at the X-ray cluster
    position. All other points were computed with the aperture
    centered on the BCG.}
  \label{fig:a1456mass}
\end{figure}
The kinematic analysis was also repeated for A~1456, where it was
found that, compared to the A~315 case, the mass estimates obtained
for A~1456 depend very little on details of the membership selection
for galaxies in the cluster field. Variations in (1) the choice of the
cluster center, (2) the selection of all vs. red-sequence galaxies
only, and (3) the choice of the limiting aperture for the analysis,
lead to variations in the virial mass estimates in the range
$4.1$--$5.2 \times 10^{14}\,h^{-1}\,M_\odot$, and in the velocity
dispersion mass estimates in the range $2.5$--$3.1 \times
10^{14}\,h^{-1}\,M_\odot$. These values agree with those listed in
\citetalias{2007A&A...461..397P} but are significantly above the mass
limit of $1.75 \times 10^{14}\,h^{-1}\,M_\odot$ derived from the
weak-lensing analysis. A comparison of the weak lensing and the
different kinematical mass estimates is shown in
Fig.~\ref{fig:a1456mass}.

\begin{figure}
  \resizebox{\hsize}{!}{\includegraphics{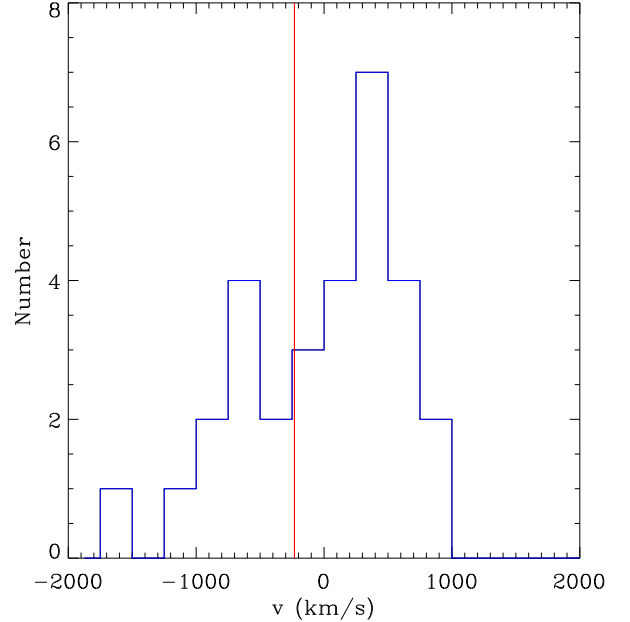}}
  \caption{Velocity histogram of A~1456. Velocities are given with
    respect to the mean velocity. The vertical red line denotes the
    velocity of the BCG.}
  \label{fig:a1456_histo}
\end{figure}
There is only marginal indication of subclustering in the A~1456 data.
There is no gradient in the radius-velocity plot, but a marginally
significant result of the \citeauthor{1988AJ.....95..985D} test for
the red-sequence galaxies ($96\%$ probability of sub-clustering).
Moreover, the velocity histogram for all galaxies within an Abell
radius ($2.15\,h^{-1}$\,Mpc) is not symmetric, with a skewness of
$-0.62$ that is marginally significant ($90$--$95\%$). Finally, the
BCG is offset from the mean velocity by $-231$\,km\,s$^{-1}$ but the
difference is within $1.6\sigma$ of the mean velocity. The velocity
histogram is shown in Fig.~\ref{fig:a1456_histo}.

\section{X-ray data analysis}
\label{sec:xray}
Based on the weak-lensing and revised kinematical mass estimates, we
also reanalyzed the X-ray data of A~315 and A~1456 in the RASS. The
cluster temperature was calculated from the cluster mass at $r_{500}$
using the mass-temperature relation in
\citet[][Table~3]{2008A&A...482..451Z}.

The X-ray luminosity, $L_\mathrm{X}$, depends on the square of the
electron density, $n^2_\mathrm{e}$, and the emission coefficient,
$\Lambda$, $L_\mathrm{X} \propto \int n^2_e \Lambda(T,Z)\mathrm{d}V$.
The luminosity can be derived from the X-ray surface brightness if the
emission coefficient is known \citep[e.g.,][]{2005A&A...429...85Z}.
The emission coefficient is a function of gas temperature and
metallicity. It can be derived in XSPEC using a spectral model for the
cluster. Here we used a combined model of ``wabs $\times$ raymond''
\citep{1983ApJ...270..119M,1977ApJS...35..419R}. The ``raymond'' model
describes the cluster thermal component including parameters of the
gas temperature, $T$, gas metallicity (assuming $Z=0.4 Z_{\odot}$),
and cluster redshift. The ``wabs'' model is set to the hydrogen column
density from the LAB survey
\citep{1997agnh.book.....H,2005A&A...440..775K}, $2.5\times
10^{20}$\,cm$^2$ for A~315 and $1.65\times 10^{20}$\,cm$^2$ for
A~1456.

Both A~315 (ID: 931706) and A~1456 (ID: 931633) were observed with the
ROSAT PSPC in the RASS. One can subtract the background from the X-ray
image to obtain the cluster image. The cluster surface brightness is
the X-ray image divided by the exposure map, which was used within the
virial radius to determine the X-ray luminosity.

The cluster temperature of A~315 derived from the weak lensing mass is
$3.33$\,keV. We used the BCG position to determine the X-ray
luminosity. An attempt to determine an X-ray position for A~315
resulted in a non-significant detection at (02:10:07.263,
$-$00:59:49.98). The X-ray luminosity changes by less than 1\% if the
cluster center is shifted to this location. The derived cluster X-ray
luminosity is $1.17\pm 0.48 \times 10^{43}\,h^{-2}$\,ergs\,s$^{-1}$ in
the $0.1$--$2.4$\,keV band. The X-ray luminosity of the BCG
(02:10:06.458, $-$01:01:56.46) is below the background level. We
therefore used the background level to estimate an upper limit to the
X-ray luminosity of the BCG inside a radius of $1.5\arcmin$. This is
$L_\mathrm{X} < 4.58 \times 10^{42}\,h^{-2}$\,erg\,s$^{-1}$ in the
$0.1$--$2.4$\,keV band.

The cluster temperature derived from the kinematical mass
($M_{200}=2.7 \times 10^{14}\,h^{-1}\,M_{\odot}$) is $3.14$\,keV. The
derived cluster X-ray luminosity is $1.18 \pm 0.48 \times
10^{43}\,h^{-2}$\,ergs\,s$^{-1}$ in the $0.1$--$2.4$\,keV band. The
X-ray luminosity varies by not more than 1\% for the different cluster
masses derived in Sects.~\ref{sec:weak-lens-mass}
and~\ref{sec:kinematic-mass}. We note that all X-ray luminosities
given here for A~315 are derived from very marginal detections. The
dominant error in these luminosities is probably the background
subtraction and the true luminosity could be substantially lower.

The cluster temperature of A~1456 derived from the weak-lensing mass
($M_{200}=0.63^{+1.13}_{-0.80}\times 10^{14}\,h^{-1}\,M_{\odot}$) is
$1.24$\,keV. We used the X-ray cluster center (12:03:45.7, +04:15:00)
in our analysis. The $1\sigma$ error of the X-ray center position is
$3\arcmin$, i.e., about $2\sigma$ away from the BCG. The X-ray
emission at the BCG position is not significantly above the
background.

The derived cluster X-ray luminosity is $2.88 \pm 0.51 \times
10^{43}\,h^{-2}$\,ergs\,s$^{-1}$ in the $0.1$--$2.4$\,keV band. The
X-ray luminosity at the BCG position (12:03:48.700, $+$04:20:44.00)
within a radius of $1.5\arcmin$ is $0.91 \pm 0.91 \times
10^{42}\,h^{-2}$\,erg\,s$^{-1}$ in the $0.1$--$2.4$\,keV band, which
is the sum of the cluster luminosity in this region and the BCG
luminosity. 

Because the weak-lensing mass estimate is compatible with no mass
being present at the cluster location, we also used the upper limit to
the weak-lensing mass ($M_{200}=1.75 \times
10^{14}\,h^{-1}\,M_{\odot}$) to estimate an upper limit of the X-ray
luminosity. This gives a cluster temperature of $2.37$\,keV. The
derived cluster X-ray luminosity is $4.43 \pm 0.65 \times
10^{43}\,h^{-2}$\,ergs\,s$^{-1}$ in the $0.1$--$2.4$\,keV band. It is
this luminosity that we plot in Fig.~\ref{fig:m-lx-relation}.

The cluster temperature corresponding to the upper value of the
kinematical mass ($M_{200}=5.18 \times 10^{14}\,h^{-1}\,M_{\odot}$) is
$4.67$\,keV. The derived cluster X-ray luminosity is $4.32 \pm 0.66
\times 10^{43}\,h^{-2}$\,ergs\,s$^{-1}$ in the $0.1$--$2.4$\,keV band.

\section{Optical cluster luminosity and richness}
\label{sec:optical}
After establishing that A~315 indeed seems to be X-ray underluminous
for its mass and that A~1456 is is in agreement with the normal
$L_\mathrm{X}$--$M_{200}$ relation but has an unusually high
kinematical mass estimate, we now examine whether the optical
properties of these clusters are in any way exceptional. Specifically,
we determined their cluster luminosities and richnesses and compared
them to the full RASS-SDSS cluster catalog.

The total optical luminosity of a cluster has to be computed after the
subtraction of the foreground and background galaxy contamination. We
considered two different approaches to the statistical subtraction of
the galaxy background. We computed the local background number counts
in an annulus around the cluster and a global background number counts
from the mean of the magnitude number counts determined in five
different SDSS sky regions, randomly chosen, each with an area of 30
deg$^2$. In our analysis, we show the results obtained using the
optical luminosity estimated with the second method. The optical
luminosity is then computed following the prescription of
\citet{2004A&A...423..449P}. The reader is referred to that paper for
a detailed discussion of the comparison between optical
luminosities calculated with different methods. To avoid selection
effects due to the slightly different redshifts of the clusters, the
optical luminosity was calculated in the same absolute magnitude
range for all the clusters. We used an absolute magnitude cut of $M_r
\leq -20$, which allowed us to sample the cluster luminosity function
down to $M^*+2$ \citep{2005A&A...433..415P}.

The cluster richness $N_\mathrm{gal}$ was calculated by summing the
background-subtracted cluster number counts used to calculate
$L_{op}$. This estimate was corrected for projection effects in the
same way as we did for $L_\mathrm{op}$, according to the prescription
given in \citet{2007A&A...464..451P}.

The cluster optical luminosity and richness were calculated within a
physical aperture of $r_{200}$. According to the weak-lensing results,
we used $r_{200}=1.30\,h^{-1}$\,Mpc for A~315. For Abell~1456 we used
the upper limit to the virial radius, $r_{200}=1.10\,h^{-1}$\,Mpc,
corresponding to the weak-lensing limit of $1.75 \times 10^{14}
M_{\odot}$.

\begin{figure} 
  \resizebox{\hsize}{!}{\includegraphics{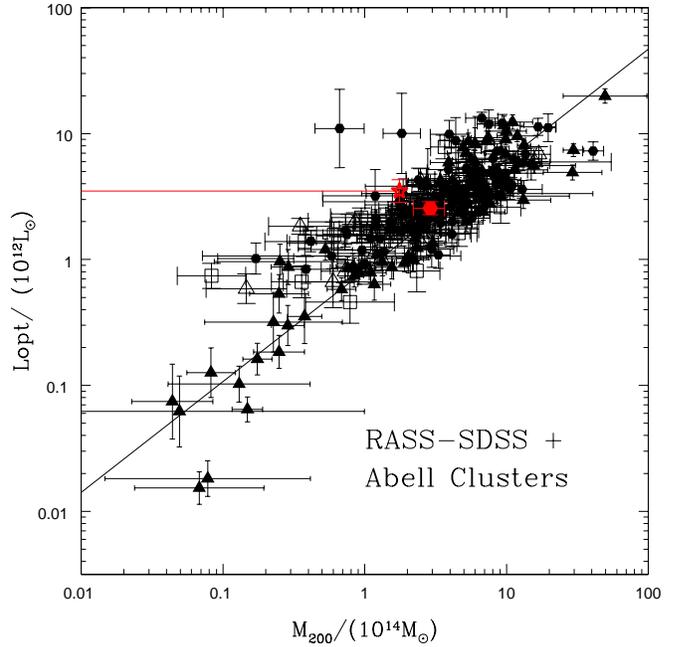}} 
  \caption{Location of Abell 315 and Abell 1456 in the
    $L_\mathrm{op}-M_{200}$ relation. The optical luminosity is calculated
    inside $r_{200}$ and is corrected for contamination due to
    projection effects. The red filled hexagon refers to Abell 315 and
    the red star is the upper limit for Abell 1456. The empty black
    squares and the filled points are the X-ray and optically selected
    clusters, respectively, used in \citet{2007A&A...464..451P}.
    The solid and dashed line are the best-fit lines obtained in
    \citet{2007A&A...464..451P}}
\label{lom} 
\end{figure}

\begin{figure} 
  \resizebox{\hsize}{!}{\includegraphics{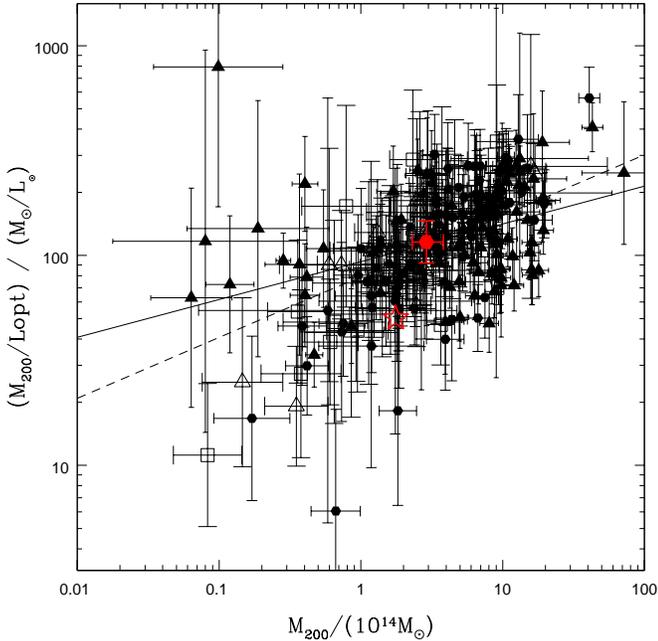}} 
  \caption{Location of Abell 315 and Abell 1456 in the
    $M_{200}/L_{op}-M_{200}$ relation. The symbols are the same as in
    Fig.~\ref{lom}}
\label{ml} 
\end{figure}

\begin{figure}
  \resizebox{\hsize}{!}{\includegraphics{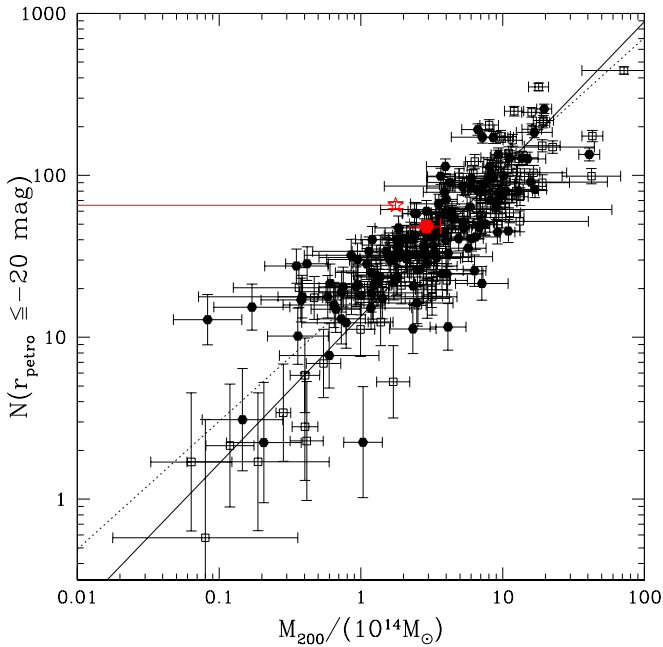}}
  \caption{Location of Abell 315 and Abell 1456 in the
    $N_{gal}-M_{200}$ relation. The symbols are the same as in
    Fig.~\ref{lom}.}
  \label{rich}
\end{figure}

Figures~\ref{lom}, \ref{ml}, and~\ref{rich} show the location of A~315
(red filled hexagon) and A~1456 (red star) in the
$L_\mathrm{opt}-M_{200}$, $M/L-M_{200}$ and $N_\mathrm{gal}-M_{200}$
relations, respectively, obtained by \citet{2007A&A...464..451P}.

\section{Discussion and conclusions}
\label{sec:discussion}
We have studied two clusters of galaxies that were previously reported
to be X-ray underluminous on the basis of their their kinematically
derived masses. To either confirm or falsify the X-ray underluminous
nature of these clusters, we obtained deep multi-color optical imaging
to measure the cluster masses using the gravitational lens effect.

Our lensing analysis of Abell~315 confirmed that this system is a
massive cluster, with a lensing mass estimate of $M_{200} =
(2.96^{+1.15\, +0.69}_{-0.75\, -0.50}) \times
10^{14}\,h^{-1}\,M_\odot$. This is substantially lower than the
kinematical mass estimate of $M_{200} = 6.61 \pm 2.71) \times
10^{14}\,h^{-1}\,M_\odot$ reported by
\citetalias{2007A&A...461..397P}. We could explain this discrepancy by
the choice of cluster center in \citetalias{2007A&A...461..397P}.
Shifting the center in the kinematical analysis to that of the BCG in
this work leads to a revised kinematical mass estimate of $M_{200} =
(2.7^{+1.1}_{-0.7}\,\pm 1.0) \times 10^{14}\,h^{-1}\,M_\odot$, which
is in excellent agreement with the weak-lensing mass estimate.

To reliably establish the X-ray underluminous nature of A~315, we also
reanalyzed the available X-ray data in the RASS. The X-ray luminosity
$L_\mathrm{X} = (1.18 \pm 0.48) \times 10^{43}\,h^{-2}$\,erg\,s$^{-1}$
determined in this work is somewhat higher than the value given by
\citet{2007A&A...461..397P} but is still significantly below the
$L_\mathrm{X}$-$M_{200}$ relation of \citet{2008MNRAS.387L..28R}.

The lensing analysis of Abell~1456 field did not show a massive
cluster. Our attempts to fit parametric models centered on the BCG
position returned only marginally significant results, as would be
expected from random fluctuations due to shape noise and the
projections of LSS. We therefore interpret the lensing measurement as
an upper limit to the mass of a cluster at this position and
redshift, $M_{200} < 1.8 \times 10^{14}\,h^{-1}\,M_\odot$. This is
significantly lower than the kinematic mass estimate that we find for
A~1456, which seems to be very robust.

A possible explanation of the much lower mass determined by weak
lensing could be that A~1456 is a bimodal cluster that the kinematic
analysis is unable to model accurately, given that indications of
subclustering are only marginal. If the cluster is indeed bimodal, as
suggested by the velocity histogram, the position of the BCG
in-between the two velocity peaks suggests that it has been displaced
from one of the two peaks by an interaction. Hence this would imply
that the kinematic mass estimate is incorrect.

The X-ray analysis of A~1456 shows a much clearer detection than in
the case of A~315. By using the upper mass limit derived from the weak
lensing analysis to determine to which radius the X-ray luminosity is
measured, we can place A~1456 right on the $L_\mathrm{X}$-$M_{200}$ relation
of \citet{2008MNRAS.387L..28R}, and the X-ray luminosity based on the
kinematic mass places A~1456 slightly above the $L_\mathrm{X}$-$M_{200}$
relation.

In summary, our results show that A~315 is indeed significantly below
the $L_\mathrm{X}$-$M_{200}$ relation, confirming that it is
underluminous compared to X-ray selected galaxy clusters of the same
size. We note that the error in the X-ray luminosity is probably
dominated by the uncertainty in the background model and higher than
the one quoted in Sect.~\ref{sec:xray}. We could not confirm that
A~1456 is an AXU cluster. The absence of a significant lensing signal
in combination with the marginal indication of bimodality in the
velocity histogram make the explanation that this is an unrelaxed
cluster merger more likely.

The revised optical luminosity and richness of both clusters are
unremarkable. The optical properties give no indication that A~315 
is either X-ray underluminous or dynamically unrelaxed. This is in
agreement with \citetalias{2007A&A...461..397P} who found that the AXU
clusters deviate from the X-ray luminosity-mass relation but are on
the optical luminosity-mass relation.

How unusual is A~315 then? \citet{2008MNRAS.387L..28R} found a scatter
in luminosity at fixed mass of $\sigma_{\ln L_\mathrm{X} | M} = 0.4$,
much higher than the scatter in the binned relation shown in
Fig.~\ref{fig:m-lx-relation}. The X-ray luminosity of A 315 given its
mass estimate would imply that the cluster deviates by $6\sigma$ from
the $L_\mathrm{X}$-mass relation, but the uncertainty in the A~315
mass estimates is rather high. However, the kinematical and
weak-lensing mass estimates are based on different physical principles
and  agree well with each other. A weighted average of the two
measurements gives a best estimate of $\bar{M}_{200} = (2.85^{+0.79\,
  +0.57}_{-0.51\, -0.45}) \times 10^{14}\,h^{-1}\,M_\odot$ , so that
the cluster deviates by $2\sigma$  from the $L_\mathrm{X}$-mass relation
along the mass axis. In all cases, A~315 appears to be an unusual cluster
compared to the population of X-ray selected clusters that define the
$L_\mathrm{X}$-mass relation.

Weak-lensing and optical-mass estimators are both affected by
projections along the line-of-sight (LOS). An extendend structure such
as a filament along the LOS or the superposition of smaller groups,
would increase both mass estimates. The X-ray luminosity, which
depends on the square of the density of the intra-cluster medium,
would not be enhanced very much in this scenario. While with present
data be used to exclude such scenarios, nothing in the data suggests
that this is a configuration that we observe. A superposition of
groups should produce kinematical substructure. With the few redshifts
available from the SDSS, this cannot be decisively excluded but we
found no evidence of substructure in Sect.~\ref{sec:kinematic-mass}. A
filament is a low density environment and as such is dominated by blue
late-type galaxies \citep{2007A&A...470..425B}. We found that 12 of
the 14 galaxies inside $r_{200}$ brighter than $L^*$ are on the
cluster red-sequence. This shows that A~315 is dominated by early-type
galaxies as one would expect for a high-density environment such as a
cluster.

Studying A~315 and similar X-ray underluminous galaxy clusters is
important to understanding the nature of the large scatter in the
cluster X-ray luminosity-mass relation. If intrinsic, as supported by
our weak-lensing mass measurement of A~315, this scatter implies that
X-ray selected cluster samples are incomplete samples in terms of
mass. An analysis of the stacked velocity histograms of all RASS-SDSS
clusters in \citetalias{2007A&A...461..397P} showed a leptokurtic
distribution, indicative of systems still in formation
\citep{2005MNRAS.361L...1W}. In such a system still in formation, the
intra-cluster gas itself may not yet have reached its final
temperature.

This scenario could be tested with further spectroscopic and X-ray
observations. Data from the SDSS do not provide enough spectra of
cluster members to show deviations from Gaussianity in the velocity
histogram of individual galaxies. A spectroscopic sample of several
hundred cluster galaxies, which can easily be obtained with modern
spectrographs, could be used to confirm the leptokurtic velocity
distribution for an individual cluster and identify infalling
substructure. Similarly, deeper X-ray observations than available in
the RASS would allow us to measure the density \emph{distribution} of
the intra-cluster gas. An improved temperature measurement due to
higher quality (background) statistics and a more self-consistent X-ray
analysis that does not rely on other methods to determine the
truncations, could significantly reduce the error in the X-ray
luminosity and provide a more definite answer to the question of how
unusual Abell~315 really is.

On the theoretical side, SPH simulations of galaxy clusters could be
used to generate mock X-ray observations, and semi-analytic models of
galaxy formation could be used to populate $N$-body halos with
galaxies. The combination of these two methods could provide further
insight into the observable properties of galaxy clusters that are not
yet fully virialized.

\acknowledgement{We are grateful to Lance Miller and Tom Kitching for
  providing their \emph{lens}fit source code to us. YYZ acknowledges
  support by the DFG through Emmy Noether Research Grant RE~1462/2 and
  by the German BMBF through the Verbundforschung under grant
  No.\,50\,OR\,0601. }

\bibliographystyle{aa}
\bibliography{AXU}

\end{document}